\documentclass[a4paper,12pt]{article}

\usepackage[english]{babel}
\usepackage[utf8x]{inputenc}
\usepackage[T1]{fontenc}
\usepackage{authblk}
\usepackage{natbib}
\usepackage{here}
\usepackage{setspace}
\usepackage{multirow}
\usepackage{lmodern}
\usepackage{amsmath, amssymb}
\usepackage{bm}
\usepackage{pdflscape}
\usepackage{authblk}
\usepackage{color}
\usepackage[table]{xcolor}

\usepackage{booktabs}
\usepackage{afterpage}
\usepackage{textcomp}
\usepackage{float}
\usepackage{enumerate}
\usepackage{rotating}

\definecolor{babyblue}{rgb}{0.54, 0.81, 0.94}
\definecolor{babypink}{rgb}{0.96, 0.76, 0.76}

\usepackage[margin=1in]{geometry}

\usepackage{amsmath}
\usepackage{graphicx}
\usepackage[colorlinks=true, allcolors=blue]{hyperref}

\usepackage{caption}
\usepackage{subcaption}

\newcommand{\pg}{P{\'o}lya-Gamma }

\definecolor{Gray}{gray}{0.85}
\newcolumntype{a}{>{\columncolor{Gray}}c}
\newcolumntype{d}{>{\columncolor{white}}c}

\begin{document}

    \begin{center}
        \vspace*{1cm}
        \large
        \textbf{Modeling complex species-environment relationships through spatially-varying coefficient occupancy models} \\
         \normalsize
           \vspace{5mm}
	    Jeffrey W. Doser\textsuperscript{1, 2}, Andrew O. Finley\textsuperscript{2, 3, 4}, Sarah P. Saunders\textsuperscript{5}, Marc K\'ery\textsuperscript{6}, Aaron S. Weed\textsuperscript{7}, Elise F. Zipkin\textsuperscript{1, 2}
         \vspace{5mm}
    \end{center}
    \small
	     \textsuperscript{1}Department of Integrative Biology, Michigan State University, East Lansing, MI, USA \\
         \textsuperscript{2}Ecology, Evolution, and Behavior Program, Michigan State University, East Lansing, MI, USA \\
         \textsuperscript{3}Department of Forestry, Michigan State University, East Lansing, MI, USA \\
         \textsuperscript{4}Department of Statistics and Probability, Michigan State University, East Lansing, MI, USA \\
         \textsuperscript{5}Science Division, National Audubon Society, New York, NY, USA \\
         \textsuperscript{6}Swiss Ornithological Institute, Sempach, Switzerland \\
         \textsuperscript{7}Northeast Temperate Inventory and Monitoring Network, National Park Service, Woodstock, VT, USA \\  
         \noindent \textbf{Corresponding Author}: Jeffrey W. Doser, email: doserjef@msu.edu; ORCID ID: 0000-0002-8950-9895 \\
         
\section*{Abstract}

Occupancy models are frequently used by ecologists to quantify spatial variation in species distributions while accounting for observational biases in the collection of detection-nondetection data. However, the common assumption that a single set of regression coefficients can adequately explain species-environment relationships is often unrealistic, especially across large spatial domains. Here we develop single-species (i.e., univariate) and multi-species (i.e., multivariate) spatially-varying coefficient (SVC) occupancy models to account for spatially-varying species-environment relationships. We employ Nearest Neighbor Gaussian Processes and \pg data augmentation in a hierarchical Bayesian framework to yield computationally efficient Gibbs samplers, which we implement in the \texttt{spOccupancy} R package. For multi-species models, we use spatial factor dimension reduction to efficiently model datasets with large numbers of species (e.g., $> 10$). The hierarchical Bayesian framework readily enables generation of posterior predictive maps of the SVCs, with fully propagated uncertainty. We apply our SVC models to quantify spatial variability in the relationships between maximum breeding season temperature and occurrence probability of 21 grassland bird species across the U.S.  Jointly modeling species generally outperformed single-species models, which all revealed substantial spatial variability in species occurrence relationships with maximum temperatures. Our models are particularly relevant for quantifying species-environment relationships using detection-nondetection data from large-scale monitoring programs, which are becoming increasingly prevalent for answering macroscale ecological questions regarding wildlife responses to global change.

\noindent \textbf{Keywords}: Bayesian, species distribution model, wildlife, monitoring, nonstationarity

\section{Introduction}

Occupancy models are a fundamental tool for ecologists and conservation practitioners to assess the probability of a site being used by a species while accounting for imperfect detection (i.e., the failure to observe a species at a site when it is truly present; \citealt{mackenzie2002, tyre2003improving}). Occupancy models typically require the collection of replicated detection-nondetection data across a set of spatial locations. Generally, multiple surveys are performed at each spatial location over a period of time in which the true presence-absence status of the species is assumed constant (i.e., the ``closure'' assumption), although such replication can also be obtained via the use of multiple observers \citep{mackenzie2005designing} or spatial sub-sampling (e.g., \citealt{sadoti2013applying}). An occupancy model consists of two sub-models: (1) an ecological process model that describes the true, partially observed presence or absence of a species as a function of covariates in a generalized linear model (GLM) framework; and (2) an observation model that specifies the observed detection-nondetection data in a similar GLM-type framework to account for false negatives, conditional on the true presence-absence of the species \citep{mackenzie2017occupancy}.

As conservation and management issues increasingly focus on multiple species, single-species occupancy models have been extended to a multi-species framework, wherein detection-nondetection data from multiple species collected via a multi-species sampling protocol are simultaneously analyzed \citep{dorazio2005, gelfand2005modelling}. Such multi-species occupancy models estimate species-specific parameters hierarchically from a shared community-level distribution, providing increased precision for species effects, as well as the ability to estimate numerous biodiversity metrics as derived parameters with fully propagated uncertainty \citep{dorazio2005}. Recent extensions of multi-species occupancy models account for residual species correlations in a joint modeling framework \citep{tobler2019joint, doser2023joint}. Together, single-species and multi-species occupancy models provide a powerful set of tools for addressing questions in ecology and conservation.

Importantly, as for many other types of species distribution models (SDMs), occupancy models are increasingly applied across broad spatial extents. This is a consequence of a growing interest in macroecological patterns, the emergence of large-scale citizen science programs (e.g., eBird, iNaturalist), and increasing availability of data from regional- to continental-scale monitoring programs. As the spatial extent of analysis increases, the common assumption of constant species-environment relationships becomes less realistic \citep{pease2022drives}. As in standard GLMs, occupancy models assume the effects of environmental covariates are constant throughout the spatial domain of the study area. Accounting for spatial variability, or nonstationarity, of the effects of environmental factors is important to accurately reflect species-environment relationships \citep{rollinson2021working} and to identify the relative effects of different environmental drivers across the range of a species  \citep{martinez2018species, sultaire2022spatial}. Failure to account for spatially-varying species-environment relationships when they exist can lead to misleading inferences and inaccurate predictions across an area of interest \citep{finley2011comparing, jarzyna2014accounting}. Such misleading inferences and predictions can lead to erroneous conclusions regarding the ecological mechanisms that determine where species occur, which could result in inappropriate management or conservation actions \citep{rollinson2021working}. 

A variety of statistical approaches exist to accommodate spatially-varying covariate effects that could be incorporated into an occupancy modeling framework. Basic approaches include: (1) stratification, in which a separate covariate effect is estimated for each individual stratum across a set of strata (e.g., ecoregions, management units; \citealt{pease2022exploring}); (2) random slopes, which is similar to stratification but instead individual stratum effects are treated as random effects from a common Gaussian distribution (e.g., \citealt{doser2021trends}); and (3) interactions, in which the effect of one environmental covariate is assumed to depend on the value of another environmental covariate (e.g., \citealt{oliver2014interactions}). While simple, such approaches are powerful for testing ecological hypotheses regarding species-environment relationships. However, these approaches restrict spatial variability as occurring only across predetermined strata or in relation to a set of known covariates. 

As species-environment relationships may vary as a result of abiotic (e.g., historical disturbance regimes) and biotic processes (i.e., density-dependent habitat selection) interacting at multiple spatial scales \citep{rollinson2021working, thorson2023spatially}, a more flexible approach is desired that can characterize complex spatially-varying species-environment relationships without pre-determined strata or without the requirement of covariates that explain spatial variation in the effects of others. Geographically weighted regression (GWR; \citealt{fotheringham2003geographically}) and generalized additive models (GAMs; \citealt{wood2006generalized}) are common approaches that provide additional flexibility to model complex spatially-varying species-environment relationships within an occupancy modeling framework. While providing more flexibility than the aforementioned approaches, GWR requires \textit{a priori} specification of parameters that control the range of spatial dependence, which can unduly impact model results and interpretation \citep{finley2011comparing, thorson2019guidance}. GAMs can accommodate spatial dependence in covariate effects via a linear combination of basis functions, but they similarly require \textit{a priori} specification of the number of knots/basis functions and can result in highly over-smoothed estimates \citep{stein2014limitations}. Bayesian spatially varying coefficient (SVC) models \citep{gelfand2003spatial} are an attractive alternative, as their hierarchical construction provides great flexibility for complex, hierarchically-structured ecological data \citep{finley2011comparing, finley2020bayesian}. In particular, SVC models are a straightforward extension of spatial GLMs that allow regression coefficients to vary smoothly across space, most commonly using some form of Gaussian process specification \citep{banerjee2014hierarchical}. While Bayesian SVC models are more complex than many of the above-mentioned models, they allow for full uncertainty propagation, do not require \textit{a priori} decisions on grid size or parameter values, are readily extensible for both single-species (i.e., univariate) and multi-species (i.e., multivariate) models, and have been shown to outperform GWR, the most commonly-used alternative, in a variety of simulation and empirical examples \citep{wheeler2007assessment, finley2011comparing}.

Here we develop computationally-efficient single-species and multi-species SVC models to allow for assessment of spatially-varying species-environment relationships while explicitly accounting for imperfect detection in an occupancy modeling framework. We employ Nearest Neighbor Gaussian Processes (NNGPs; \citealt{datta2016hierarchical}) and \pg data augmentation \citep{polson2013} in our Bayesian implementations to yield computationally efficient Gibbs samplers for both single-species and multi-species models, which we make available via new functions in the \texttt{spOccupancy} \texttt{R} package \citep{doser2022spoccupancy}. For multi-species models, we use a spatial factor model \citep{hogan2004bayesian} to ensure computational efficiency for multi-species models with a moderate to large (e.g., $> 10$) number of species while simultaneously accounting for residual correlations between species \citep{doser2023joint}. Our motivating data set for these models consists of detection-nondetection data on 21 grassland bird species from the North American Breeding Bird Survey \citep{pardieck2020north}. Grassland birds have declined precipitously in recent years \citep{rosenberg2019decline}, and thus there is considerable conservation interest in understanding environmental drivers of their occurrence \citep{stanton2018analysis}. Here, we seek to quantify the relationships between maximum breeding season temperature and occurrence of grassland bird species, with a particular focus on assessing spatial variation in this relationship among species, and how the estimated relationship relates to another driver of grassland bird occurrence, the amount of grassland habitat, at both the individual species level and on average across the community.

The remainder of this paper is organized as follows. In Section \ref{SVC-SSOMs} we describe the single-species SVC occupancy model, while in Section \ref{SVC-MSOMs} we detail the multi-species SVC occupancy model. In Sections \ref{pgSampler}-\ref{software}, we discuss additional details on the hierarchical Bayesian implementations of our models, including a brief discussion of our implementation of the SVC occupancy models in the \texttt{spOccupancy} \texttt{R} package. In Section \ref{simulation}, we perform a simulation study as a proof of concept to illustrate the pitfalls of ignoring spatial variability in species-environment relationships when present. In Section \ref{caseStudy}, we apply our models to the grassland bird data set and compare their performance with a series of alternative models. Finally, in Section \ref{discussion} we discuss the broader applicability of our proposed models for modeling species distributions and their environmental drivers.

\section{Models}

\subsection{Single-species spatially-varying coefficient occupancy models}\label{SVC-SSOMs}

\subsubsection{Process model}

Let $\bm{s}_j$ denote the spatial coordinates of site $j$, where $j = 1, \dots, J$. We define $z(\bm{s}_j)$ as the true presence (1) or absence (0) of the target species at site $j$ with spatial coordinates $\bm{s}_j$. We model $z(\bm{s}_j)$ as 
\begin{equation}\label{z}
    z(\bm{s}_j) \sim \text{Bernoulli}(\psi(\bm{s}_j)), 
\end{equation}

where $\psi(\bm{s}_j)$ is the occupancy probability of the species at site $j$. We model $\psi(\bm{s}_j)$ according to 
\begin{equation}\label{psi}
   \text{logit}(\psi(\bm{s}_j)) = (\beta_1 + \delta_1\text{w}_1(\bm{s}_j)) + \sum_{h = 2}^H\text{x}_h(\bm{s}_j)\{\beta_h + \delta_h\text{w}_h(\bm{s}_j)\},
\end{equation}

where $\beta_1$ is an intercept, $\text{x}_h(\bm{s}_j)$ is the $h$th covariate with $h = 2, \dots, H$, $\beta_h$ is the non-spatial effect of covariate $\text{x}_h(\bm{s}_j)$, and $\text{w}_1(\bm{s}_j)$ and $\text{w}_h(\bm{s}_j)$ are spatially-varying effects for the intercept and covariates, respectively. We use indicator variables $\delta_h$ for $h = 1, \dots, H$ to indicate those covariates whose effects vary spatially ($\delta_h = 1$) and those whose effects are assumed constant ($\delta_h = 0$). Note, the model reduces to a traditional single-species occupancy model when $\delta_h = 0$ for all $h$ and to a spatial occupancy model \citep{johnson2013spatial, doser2022spoccupancy} when $\delta_1 = 1$ and $\delta_h = 0$ for all $h > 1$. For later use, define $\tilde{H}$ as the total number of spatially-varying effects estimated in the model (i.e., $\tilde{H} = \sum_{h = 1}^H\delta_h$), define $\tilde{\textbf{x}}(\bm{s}_j)$ as the $\tilde{H} \times 1$ vector of covariates at location $j$ (including an intercept if applicable), and define $\tilde{\beta}_h(\bm{s}_j) = \beta_h  + \text{w}_h(\bm{s}_j)$ as the spatially-varying coefficients for those effects with $\delta_h = 1$.

Let $\mathcal{L} = \{\bm{s}_1, \bm{s}_2, \dots, \bm{s}_J\}$ be the set of sampled spatial locations, and define $\textbf{w}_h$ as a $J \times 1$ vector of the spatial random effects for covariate $h$ for each of the $\tilde{H}$ effects with corresponding $\delta_h = 1$. The spatially-varying effects $\textbf{w}_h$ serve as local adjustments of the covariate effects (or intercept) at each site $j$ from the overall non-spatial effect $\beta_h$, resulting in the covariate having a unique effect (i.e., $\tilde{\beta}_h(\bm{s}_j))$ on species occupancy probability at each site $j$. Following \cite{gelfand2003spatial}, we envision each $h = 1, \dots, \tilde{H}$ spatially-varying effects $\text{w}_h(\bm{s}_j)$ as realizations of a smooth latent surface $\{\text{w}_h(\bm{s}) \mid \bm{s} \in \mathcal{D}\}$, where $\mathcal{D}$ is the geographical domain of interest. Following standard approaches for modeling species distributions (e.g., \citealt{latimer2006building}), we use Gaussian Processes (GPs) to model each of the $\tilde{H}$ smooth functions across the spatial domain. By definition, a GP model for the $h$th spatially-varying surface $\{\text{w}_h(\bm{s})\}$ implies that for any finite set of locations $\mathcal{L}$, the vector of random effects $\textbf{w}_h$ follows a zero-mean multivariate Gaussian distribution with a $J \times J$ covariance matrix $\bm{C}_h(\bm{s}, \bm{s}', \bm{\theta}_h)$ that is a function of the distances between any pair of site coordinates $\bm{s}$ and $\bm{s}'$ and a set of parameters ($\bm{\theta}_h$) that govern the spatial process according to a parametric covariance function. In our subsequent simulations and case study, we use an exponential covariance function such that $\bm{\theta}_h = \{\sigma^2_h, \phi_h\}$, where $\sigma^2_h$ is a spatial variance parameter and $\phi_h$ is a spatial decay parameter. Large values of $\sigma^2_h$ indicate large variation in the magnitude of a covariate effect across space, while values of $\sigma^2_h$ close to 0 suggest little spatial variability in the magnitude of the effect. $\phi_h$ controls the distance-dependent decay of the spatial dependence in the covariate effect and is inversely related to the spatial range, such that when $\phi_h$ is small, the covariate effect has a larger range of spatial dependence and varies more smoothly across space compared to larger values of $\phi_h$. When using an exponential correlation function, the effective spatial range, or the distance at which the spatial correlation between points drops to 0.05 \citep{banerjee2014hierarchical}, corresponds to approximately 3 / $\phi$ (i.e., since $3 \approx -\text{log}(0.05))$.

Both frequentist and Bayesian estimation of the model defined by (\ref{z}) and (\ref{psi}) requires taking the inverse and determinant of $\tilde{H}$ dense $J \times J$ covariance matrices (i.e., $\bm{C}_t(\bm{s}, \bm{s}', \bm{\theta}_h)$) that involves $O(J^3)$ computations for each of the $\tilde{H}$ spatially-varying coefficients (floating point operations or FLOPs), which quickly renders such an approach impractical for even moderately sized data sets (i.e., hundreds of spatial locations). Here we replace the GP prior for the spatially-varying coefficients with a Nearest Neighbor Gaussian Process (NNGP) prior \citep{datta2016hierarchical}. The NNGP is a valid GP that is based on writing the full multivariate Gaussian distribution for $\textbf{w}_h$ as a product of conditional densities, such that 
\begin{equation}\label{wConditional}
   p(\textbf{w}_h) = p(\text{w}_h(\bm{s}_1)) \cdot p(\text{w}_h(\bm{s}_2) \mid \text{w}_h(\bm{s}_1)) \cdots p(\text{w}_h(\bm{s}_J) \mid \text{w}_h(\bm{s}_{J - 1}), \dots, \text{w}_h(\bm{s}_1)),
\end{equation}

where $p(\cdot)$ denotes a probability density function. The NNGP prior achieves computational efficiency by replacing the conditioning sets on the right-hand side of (\ref{wConditional}) with a set of new conditioning sets, whose maximum size is determined by a pre-specified number of neighbors, $m$, where $m << J$. \cite{datta2016hierarchical} showed that $m = 15$ provides nearly identical inference to the full GP under a variety of scenarios. Let $N(\bm{s}_j)$ denote the set of at most $m$ neighbors for location $\bm{s}_j$. Following \cite{vecchia1988estimation}, we set $N(\bm{s}_j)$ to be the set of at most $m$ nearest neighbors of $\bm{s}_j$ from $\{\bm{s}_1, \bm{s}_2, \dots, \bm{s}_{j - 1}\}$ with respect to Euclidean distance. Note, this requires the set of $\mathcal{L}$ locations to have some prespecified ordering, where here we order the coordinates along the horizontal axis. Through careful construction of the neighbor sets and set of spatial locations as a directed acyclic graph, Gaussian distribution theory reveals the NNGP prior yields a new joint density for $\textbf{w}_h$, denoted $\tilde{p}(\textbf{w}_h)$. Let $\textbf{w}_h(N(\bm{s}_j))$ denote the at most $m$ realizations of the $h$th NNGP at the locations in the neighbor set $N(\bm{s}_j)$. Let $C(\cdot, \bm{\theta}_h)$ denote the covariance function of the original Gaussian Process (GP) from which the $h$th NNGP is derived. For any two sets $A_1$ and $A_2$, define $\text{C}_{A_1, A_2}(\bm{\theta}_h)$ as the covariance matrix between the observations in $A_1$ and $A_2$ for the $h$th GP.  For all $h = 1, \dots, \tilde{H}$, our NNGP prior for $\textbf{w}_h$ thus takes the form
\begin{equation}\label{NNGP}
     \tilde{p}(\textbf{w}_h) = \prod_{j = 1}^J \text{Normal}(\text{w}_h(\bm{s}_j) \mid \textbf{w}_h(N(\bm{s}_j))\textbf{b}_h(\bm{s}_j), \text{f}_h(\bm{s}_j)), 
\end{equation}
where $\textbf{b}_h(\bm{s}_j)$ is defined as 
\begin{equation}\label{bNNGP}
  \textbf{b}_h(\bm{s}_j) = \textbf{C}_{\bm{s}_j, N(\bm{s}_j)}(\bm{\theta}_h)\textbf{C}^{-1}_{N(\bm{s}_j), N(\bm{s}_j)}(\bm{\theta}_h), 
\end{equation}
with $\textbf{b}_h(\bm{s}_1) = \textbf{0}$, and $\text{f}_h(\bm{s}_j)$ is defined as 
\begin{equation}\label{fNNGP}
  \text{f}_h(\bm{s}_j) = \textbf{C}_{\bm{s}_j, \bm{s}_j}(\bm{\theta}_h) - \textbf{C}_{\bm{s}_j, N(\bm{s}_j)}(\bm{\theta}_h)\textbf{C}^{-1}_{N(\bm{s}_j), N(\bm{s}_j)}(\bm{\theta}_h)\textbf{C}_{N(\bm{s}_j), \bm{s}_j}(\bm{\theta}_h). 
\end{equation}

\subsubsection{Observation model}

To account for imperfect detection in an occupancy modeling framework, $k = 1, \dots, K(\bm{s}_j)$ sampling replicates are obtained at each site $j$ to estimate whether a nondetection of the target species is truly an absence \citep{mackenzie2002, tyre2003improving}. We model the observed detection (1) or nondetection (0) of a study species at site $j$, denoted $y_k(\bm{s}_j)$, conditional on the true latent occupancy process $z(\bm{s}_j)$, following
\begin{equation}\label{yDet}
        y_{k}(\bm{s}_j) \mid z(\bm{s}_j) \sim \text{Bernoulli}(p_{k}(\bm{s}_j)z(\bm{s}_j)),
\end{equation}

where $p_k(\bm{s}_j)$ is the probability of detecting the species at site $j$ during replicate $k$ given the species is truly present at the site. We model detection probability as a function of site and/or observation-level covariates according to 
\begin{equation}\label{pDet}
  \text{logit}(p_{k}(\bm{s}_j)) = \textbf{v}_{k}(\bm{s}_j)^\top\bm{\alpha}, 
\end{equation}

where $\bm{\alpha}$ is a vector of regression coefficients (including an intercept) that describe the effect of site and/or observation covariates $\textbf{v}_{k}(\bm{s}_j)$ on detection. Note that, following the standard occupancy model, we assume independence between the replicate surveys, conditional on the true occupancy status $z(\bm{s}_j)$ and covariates \textbf{v}$_k(\bm{s}_j)$, the true occupancy status does not change over the $K(\bm{s}_j)$ replicate surveys, and there are no false positives (i.e., if $z(\bm{s}_j) = 0$, then $y_k(\bm{s}_j) = 0$ for all $k$). We assume effects of covariates on detection probability are constant over space, but in principle spatially-varying covariate effects could be added to the detection model as well, using the same process described above.

\subsubsection{Prior specifications}\label{priorsSSOM}

We assign independent Gaussian priors to all non-spatial regression coefficients ($\bm{\beta}$ and $\bm{\alpha}$), independent inverse-Gamma priors to the spatial variance parameters ($\sigma^2_h$) for each spatially-varying effect, and independent uniform priors for the spatial decay parameters ($\phi_h$). 

\subsection{Multi-species spatially-varying coefficient occupancy models}\label{SVC-MSOMs}

Now consider the case where there are multiple species of interest, $N$, that are observed during data collection. We seek to jointly model the occupancy of the $N$ species in a single model that accommodates residual correlations between species and allows for sharing of information across species via random effects. Such multi-species approaches often yield improved precision and accuracy of estimates compared to single-species models \citep{clark2014more, zipkin2010multi}. Using similar notation to that for the single-species models, we model the true presence-absence state of species $i$ at site $\bm{s}_j$ following
\begin{equation}\label{zMS}
   z_i(\bm{s}_j) \sim \text{Bernoulli}(\psi_i(\bm{s}_j) z^\ast_i(\bm{s}_j)), 
\end{equation}

where $\psi_i(\bm{s}_j)$ is the occupancy probability of species $i$ at site $j$, and $z^\ast_i(\bm{s}_j)$ is a binary auxiliary data source indicating whether site $j$ is within the known range of species $i$. Such data can be obtained from a variety of sources, including international databases (e.g., BirdLife International, IUCN), field guides, or expert opinion. We suggest buffering the auxiliary data range map by a suitable distance to account for potential inaccuracies in these auxiliary data. Inclusion of such auxiliary range data can drastically reduce the computational burden of the model if certain species can only exist at a subset of the spatial locations in $\mathcal{L}$ \citep{socolar2022biogeographic}. If auxiliary range data are not available, $z^\ast_i(\bm{s}_j)$ can be removed from (\ref{zMS}) (or equivalently, $z^\ast_i(\bm{s}_j) = 1$ for all $j$). 

At sites within a species' range, we model species-specific occupancy probability as 
\begin{equation}\label{psiMS}
   \text{logit}(\psi_i(\bm{s}_j)) = (\beta_{i, 1} + \delta_1\text{w}^\ast_{i, 1}(\bm{s}_j)) + \sum_{h = 2}^H\text{x}_h(\bm{s}_j)\{\beta_{i, h} + \delta_h\text{w}^\ast_{i, h}(\bm{s}_j)\},
\end{equation}

with all parameters as defined before, but now spatial and non-spatial effects are unique to each species. We assume the same variables have spatially-varying effects (or not) for all species (i.e., $\delta_{i, h} = \delta_h$ for all $i$), although this could be modified to allow $\delta_h$ to vary by species by either setting values \textit{a priori} or estimating the indicator variables within the model itself. We model the non-spatial component of the $h$th regression coefficient for each species $i$ hierarchically from a common community-level distribution to share information across species \citep{dorazio2005, gelfand2005modelling}. More specifically, we have 
\begin{equation}\label{betaMS}
     \beta_{i, h} \sim \text{Normal}(\mu_{\beta_h}, \tau^2_{\beta_h}), 
\end{equation}
where $\mu_{\beta_h}$ is the average non-spatial effect across all species in the community, and $\tau^2_{\beta_h}$ is the variability in the non-spatial effect across all species.

We seek to jointly model the species-specific spatial effects, $\textbf{w}^\ast_{i, h}$, to account for correlation in species-specific responses to covariates, as well as residual correlations between species after accounting for their relationships with any covariates included in the model. For a small number of species (e.g., 5), a linear model of coregionalization (LMC) framework \citep{gelfand2004} is a viable solution, but such an approach quickly becomes computationally intractable as the number of species in the community increases (e.g., $> 5$). Instead, we use a spatial factor model \citep{hogan2004bayesian}, a dimension reduction technique that accounts for correlations in species-specific responses, while drastically reducing computational run time compared to a LMC that requires estimation of a full $N \times N$ cross-covariance matrix for each effect that is assumed to vary spatially. Here, we decompose $\text{w}^\ast_{i, h}(\bm{s}_j)$ into a linear combination of $q$ latent factors and their associated species-specific coefficients (i.e., factor loadings). Thus for each SVC in the model, we have
\begin{equation}\label{wStar}
     \text{w}^\ast_{i, h}(\bm{s}_j) = \bm{\lambda}^\top_{i, h}\textbf{w}_{h}(\bm{s}_j),
\end{equation}
where $\bm{\lambda}_{i, h}^\top$ is the $i$th row of factor loadings from the $N \times q$ loadings matrix $\bm{\Lambda}_h$, and $\textbf{w}_h(\bm{s}_j)$ is a $q \times 1$ vector of independent spatial factors at site $j$. As in the single-species SVC occupancy model, we model each of the $r = 1, \dots, q$ spatial factors for each of the $\tilde{H}$ spatially-varying effects with an NNGP prior following
\begin{equation}\label{NNGPMS}
     \tilde{p}(\textbf{w}_{r, h}) = \prod_{j = 1}^J \text{Normal}(\text{w}_{r, h}(\bm{s}_j) \mid \textbf{w}_{r, h}(N(\bm{s}_j))\textbf{b}_{r, h}(\bm{s}_j), \text{f}_{r, h}(\bm{s}_j)), 
\end{equation}
with $\textbf{b}_{r, h}(\bm{s}_j)$ and $\text{f}_{r, h}(\bm{s}_j)$ defined in (\ref{bNNGP}) and (\ref{fNNGP}), respectively.  

For each SVC, we can derive an inter-species covariance matrix $\bm{\Sigma}_h = \bm{\Lambda}_h\bm{\Lambda}_h^\top$, which has rank $q << N$, and, thus, is singular. However, the inter-species covariance matrices can still be used to detect species clustering \citep{shirota2019sinica, doser2023joint}. For a spatially-varying intercept, the inter-species covariance matrix provides information on the residual co-occurrence patterns between each pair of species across space after accounting for any covariates included in the model. This can be used to generate hypotheses about the abiotic and/or biotic drivers of residual species co-occurrence patterns (e.g., \citealt{tobler2019joint}). For a spatially-varying covariate effect, a positive correlation between species in $\bm{\Sigma}_h$ indicates similar responses to that environmental covariate across space. One possible application of this is as a model-based ordination technique to identify groups of species that respond similarly to environmental variables \citep{thorson2015spatial}.

Analogous to the single-species case, we model the observed detection-nondetection of each species $i$ at site $j$ during replicate survey $k$, $y_{i, k}(\bm{s}_j)$ conditional on the true presence-absence of each species, $z_{i}(\bm{s}_j)$, following (\ref{yDet}) and (\ref{pDet}), with all parameters now varying by species. We model the species-specific detection regression coefficients ($\bm{\alpha}_i$) hierarchically, analogous to the non-spatial occupancy regression coefficients in (\ref{betaMS}).

\subsubsection{Prior specification and identifiability considerations}

We assume Gaussian priors for all mean parameters and inverse-Gamma priors for variance parameters. Additional restrictions on the factors loadings matrix $\bm{\Lambda}_h$ for each spatially-varying coefficient $h$ are required to ensure identifiability \citep{taylor2019spatial}. We fix all elements in the upper triangle to 0 and set the diagonal elements to 1. We additionally fix the spatial variance parameters $\sigma^2_h$ of each latent spatial process to 1. We assign standard Gaussian priors for the lower triangular elements in $\bm{\Lambda}$ and assign each spatial range parameter $\phi_{r, h}$ an independent uniform prior.

\subsection{Implementation via a \pg Gibbs sampler}\label{pgSampler}

We implement both single-species and multi-species SVC occupancy models in a hierarchical Bayesian framework using Markov chain Monte Carlo (MCMC), which we briefly describe below in the context of multi-species models with additional details in Supplemental Information S1. Using standard approaches, the full conditional distributions for $\bm{\beta}_i$, $\bm{\alpha}_i$, and $\textbf{w}_{r, h}(\bm{s})$ are not available in closed form, resulting in the use of less efficient samplers for these parameters, e.g., a Metropolis-Hastings random walk. Instead, we implement an efficient \pg data augmentation scheme \citep{polson2013}, which yields fully Gibbs updates for all parameters except $\bm{\phi}$ in the proposed models. Briefly, for each species $i$ we introduce two sets of \pg auxiliary variables, $\bm{\omega}_{i, \psi} = (\omega_{i, \psi}(\bm{s}_1), \dots, \omega_{i, \psi}(\bm{s}_J))$ and $\bm{\omega}_{i, p} = (\omega_{i, p, 1}(\bm{s}_1), \dots, \omega_{i, p, K(\bm{s}_{J})}(\bm{s}_J))$, for the occupancy (i.e., process) and detection (i.e., observation) sub-models, respectively. Note, for the detection sub-model, augmented data points are required for each replicate survey at each spatial location, although gains in efficiency are possible by only sampling the auxiliary variables at sites where $z_i(\bm{s}_j) = 1$ for any given MCMC iteration (see Supplemental Information S1). Both $\bm{\omega}_{i, \psi}$ and $\bm{\omega}_{i, p}$ are assigned \pg priors with shape parameter 1 and tilting parameter 0 (i.e., $\text{PG}(1, 0)$). Continuing with the occupancy process, we can re-express the Bernoulli model in (\ref{z}) as 
\begin{equation}\label{pgOmega}
\begin{split}
\psi_i(\bm{s}_j)^{z_i(\bm{s}_j)} (1 - \psi_i(\bm{s}_j))^{1 - z_i(\bm{s}_j)} &= \frac{\text{exp}(\textbf{x}(\bm{s}_j)^{\top}\bm{\beta}_i + \tilde{\textbf{x}}(\bm{s}_j)^\top\textbf{w}_i^\ast(\bm{s}_j))^{z_i(\bm{s}_j)}}{1 + \text{exp}(\textbf{x}^{\top}_j\bm{\beta}_i + \tilde{\textbf{x}}(\bm{s}_j)^\top\textbf{w}_i^\ast(\bm{s}_j))} \\
&= \text{exp}(\kappa_i(\bm{s}_j)(\textbf{x}(\bm{s}_j)^\top\bm{\beta}_i + \tilde{\textbf{x}}(\bm{s}_j)^\top\textbf{w}_i^\ast(\bm{s}_j)) ) \times \\
&\hspace{5mm} \int \text{exp}(-\frac{\omega_{i, \psi}(\bm{s}_j)}{2}(\textbf{x}(\bm{s}_j)^{\top}\bm{\beta}_i + \tilde{\textbf{x}}(\bm{s}_j)^\top\textbf{w}_i^\ast(\bm{s}_j)))^2 \times \\
&\hspace{10mm}p(\omega_{i, \psi}(\bm{s}_j) \mid 1, 0) d\omega_{i, \psi}(\bm{s}_j),
\end{split}
\end{equation}

where $\kappa_i(\bm{s}_j) = z_i(\bm{s}_j) - 0.5$ and $p(\omega_{i, \psi}(\bm{s}_j))$ is the probability density function of a \pg distribution with shape 1 and tilting parameter 0 \citep{polson2013}. This re-expression of the Bernoulli model results in Gibbs updates for both the non-spatial ($\bm{\beta}_i$) and spatial ($\textbf{w}_{r, h}(\bm{s})$) effects. An analogous re-expression is used for the likelihood in (\ref{yDet}) to yield Gibbs updates for the detection coefficients $\bm{\alpha}_i$.

\subsection{Posterior predictive inference}

The hierarchical Bayesian approach readily enables posterior predictive inference at any new location, $\bm{s}_0$, with full uncertainty propagation. Prediction of occupancy probability across a region of interest (i.e., a species distribution map) proceeds in two steps. For multi-species models and for each posterior sample, $l$, we generate estimates of each spatial factor, $\textbf{w}_{r, h}$, from 
\begin{equation}\label{wPrediction}
    \text{w}^{(l)}_{r, h}(\bm{s}_0) \mid \cdot \sim \text{Normal}(\textbf{w}^{(l)}_{r, h}(N(\bm{s}_0))\textbf{b}^{(l)}_{r, h}(\bm{s}_0), \text{f}^{(l)}_{r, h}(\bm{s}_0)).
\end{equation}

Estimates from (\ref{wPrediction}) for the $q$ spatial factors can be multiplied by the factor loadings of a given species (i.e., $\bm{\lambda}_{i, h}^{(l)}$) to generate $\text{w}_{i, h}^{\ast, (l)}(\bm{s}_0)$ as in (\ref{wStar}), which can subsequently be added to estimates of the non-spatial component of the covariate effect (i.e., $\beta^{(l)}_{i, h}$) obtained when fitting the model to generate posterior predictive estimates of the full SVC at new locations (i.e., $\tilde{\beta}^{(l)}_{i, h}(\bm{s}_0)$). Conditional on estimates of $\text{w}_{i, h}^{\ast, (l)}(\bm{s}_0)$, we can then generate predictions of occupancy probability and the true presence-absence state of species $i$ at new locations following
\begin{equation}\label{psiPrediction}
    \begin{split}
        \text{logit}(\psi_i^{(l)}(\bm{s}_0)) &= (\beta_{i, 1}^{(l)} + \delta_1\text{w}^{\ast, (l)}_{i, 1}(\bm{s}_0)) + \sum_{h = 2}^H\text{x}_h(\bm{s}_0)\{\beta_{i, h}^{(l)} + \delta_h\text{w}_{i, h}^{\ast, (l)}(\bm{s}_0)\}, \\
        z_i^{(l)}(\bm{s}_0) &\sim \text{Bernoulli}(\psi_i^{(l)}(\bm{s}_0)).
    \end{split}
\end{equation}

\subsection{Software Implementation}\label{software}

We implement single-species and multi-species SVC occupancy models in the functions \texttt{svcPGOcc} and \texttt{svcMsPGOcc}, respectively, in the \texttt{spOccupancy} R package \citep{doser2022spoccupancy} to facilitate the use of SVC occupancy models in user-friendly software. A detailed vignette that walks through the models and shows how to fit them in \texttt{spOccupancy} is freely available on the package website (\url{https://www.jeffdoser.com/files/spoccupancy-web/articles/index.html}).

\section{Simulation study} \label{simulation}

\subsection{Methods}

As a proof of concept, we used simulation to assess how failing to account for spatially-varying species-environment relationships may affect inference for various levels of spatial dependence in the covariate effect (e.g., no spatial variability, small vs. high variability, short vs. long range spatial dependence). We simulated data for a single species from $J = 400$ sites across a unit square and $K = 5$ replicate surveys at each site for use in a single-species occupancy modeling framework, where detection probability was set to moderate (average = 0.45) and varying positively (0.4 on the logit scale) with a standardized observation-level covariate. We simulated species occupancy as a function of an intercept (logit-scale $\beta_0 = 0$) and a single covariate (logit-scale $\beta_1 = 0$), whose effects both varied across space following (\ref{psi}) and (\ref{NNGP}) using an exponential correlation function. For the intercept, we set the spatial variance to 1 and the spatial decay parameter to $\phi = 3 / 0.8$. For the SVC of the covariate effect, we varied the spatial variance parameter across four values ($\sigma^2 = \{0.1, 0.5, 1, 2\}$) and the spatial decay parameter across five values ($\phi = \{3 / 0.1, 3 / 0.5, 3 / 0.8, 3 / 3, 3 / 100\}$) to assess how failing to account for spatial variability in the covariate effect is related to the spatial characteristics of the effect. See Supplemental Information S2 Figure S1 for the resulting spatial pattern in the SVC under all 20 combinations of these parameter values. Since we simulated the data set across a unit square, the values for $\phi$ correspond to very fine-scale spatial variability in the covariate effect (i.e., 3 / 0.1), more broad-scale spatial variability in the covariate effect relative to the simulated study area (i.e., 3 / .5, 3/ .8, 3 / 3), and extremely broad-scale spatial variability in the effect (i.e., 3 / 100) such that the effect is essentially constant within the simulated study region. We also generated data where the covariate effect was constant to assess performance of the SVC occupancy model when no spatial variability was present. We simulated 50 data sets for each combination of parameter values, resulting in 1050 simulated data sets. We compared the performance of three models: (1) a basic occupancy model with no spatially-varying intercept or covariate effects; (2) a spatial occupancy model with a spatially-varying intercept but no spatially-varying covariate effects; and (3) the full SVC occupancy model. We used the Widely Applicable Information Criterion (WAIC) as a measure of model fit \citep{watanabe2010} and four-fold cross-validation using model deviance as a metric of predictive performance. 

\subsection{Results}

The benefits of modeling spatial variability in SVC occupancy models were dependent on the characteristics of the spatial dependence in the covariate effect (Table \ref{tab:simStudy1}). Improvements in model performance of the SVC occupancy model were highest for covariate effects with a large spatial variance and an effective spatial range of 50\% or 80\% of the area according to both WAIC and four-fold cross-validation deviance (Table \ref{tab:simStudy1}, Appendix S2: Table S2). As expected, when the spatial variance in the covariate effect was small (0.1, 0.5), an SVC occupancy model only yielded minor improvements in WAIC compared to a spatial occupancy model that assumed a stationary covariate effect, and either no improvements or very small improvements in cross-validation deviance. Predictive performance of SVC models was generally worse or only marginally improved when the effective range of the covariate effect was small (10\% of study area), regardless of the spatial variance. As the effective spatial range decreases towards zero, the spatial correlation in the SVCs occurs over a smaller distance, and thus the estimate of the SVC at any given location is informed by a smaller number of data points. For binary data, it is not possible to estimate an unstructured site-level random effect \citep{bolkerGLMM}, which likely contributed to why we saw negligible differences in predictive performance between the SVC occupancy model and the spatial occupancy model when the effective spatial range was small. When the effective spatial range of the SVC was far larger than the study area (i.e., $\phi = 3 / 100$), the covariate effect was essentially constant within the simulated study area (Supplemental Information S2 Figure S1) and the SVC model did not show any improvements over the occupancy model with a spatially-varying intercept and constant covariate effect. This suggests that an SVC occupancy model will only provide improvements if the spatial extent of the study region is moderately large relative to the spatial variation in the covariate effect. Not surprisingly, when the simulated covariate effect was assumed constant, the SVC model showed no improvements over the spatial occupancy model with a constant covariate effect. Overall, this simulation study shows that SVC occupancy models outperform models with a constant covariate effect when the effect varies across space, but the benefits of SVC models are dependent on the range of the spatial variation in the effect relative to the size of the study area.

\section{Case study} \label{caseStudy}

\subsection{Methods}

Our motivation for the aforementioned model and software development stems from a desire to quantify spatial variability in the effects of climate and land cover on grassland bird communities across the US. Identifying spatial variation in species-environment relationships has critical implications for conservation and management efforts, such as species re-introductions and habitat restoration \citep{saunders2022unraveling}. Here, our objective is to quantify the spatially-varying relationships between maximum (i.e., extreme) breeding season temperature and occurrences of 21 grassland bird species across the continental US. We hypothesize the effect of maximum temperature on occurrence will vary spatially as a result of non-linear species-specific thermal tolerances, spatial variation in habitat availability (i.e., percent grassland cover; \citealt{zuckerberg2018effects}), and additional interactions with unknown abiotic and biotic factors.

\begin{figure}
    \centering
    \includegraphics[width=9cm]{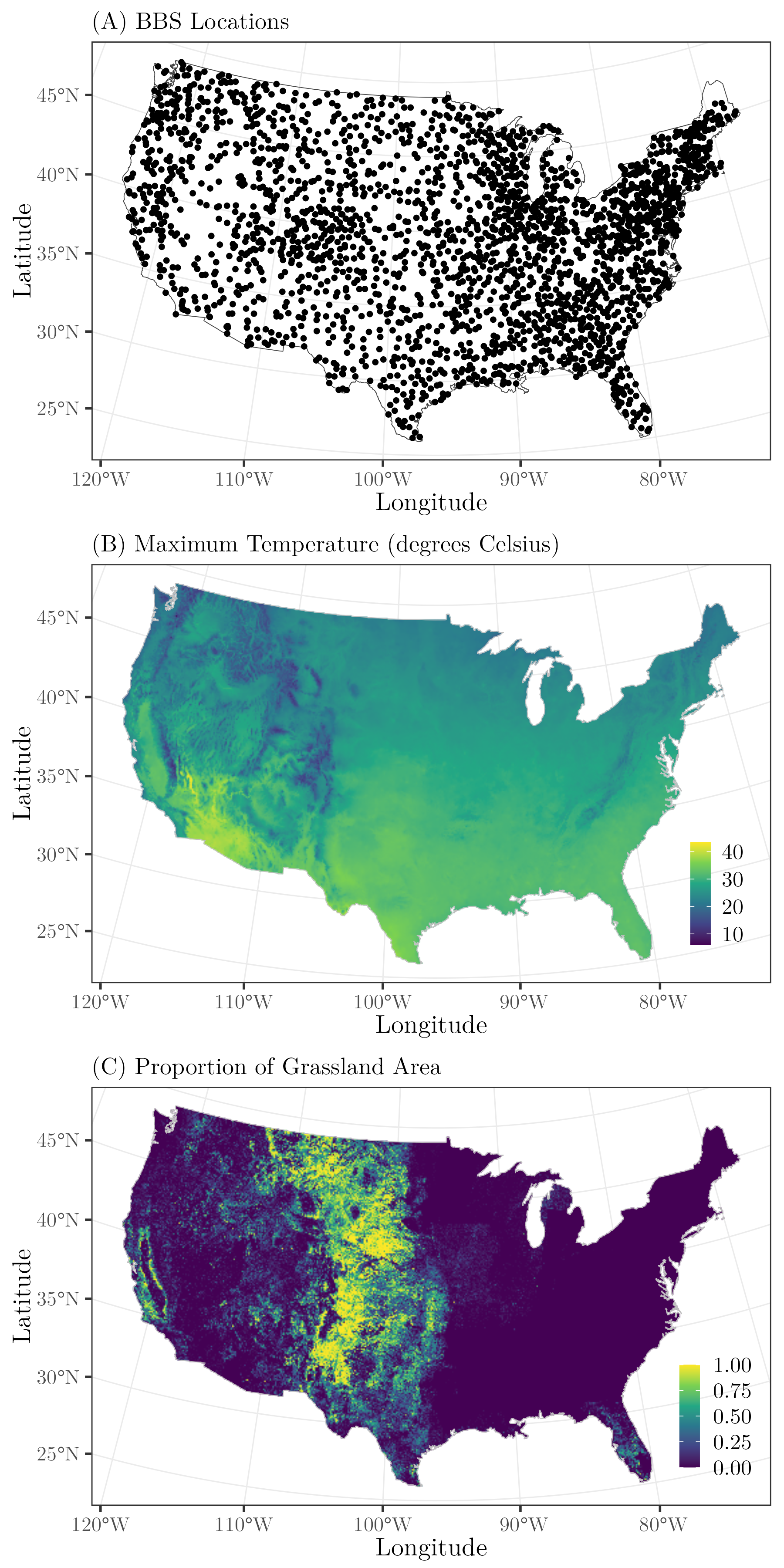}
    \caption{Spatial distribution of the $J = 2,486$ North American Breeding Bird Survey (BBS) routes used in the grassland bird case study (A) and the two covariates, maximum breeding season temperature (B) and proportion of grassland area (C), included in the model.}
    \label{fig:studyLocations}
\end{figure}

We used bird detection-nondetection data across $J = 2,486$ locations in the continental US, sampled in 2019 as part of the North American Breeding Bird Survey (BBS; \citealt{pardieck2020north}; Fig. \ref{fig:studyLocations}A). The BBS is a volunteer-based program where observers perform point count surveys at 50 points (i.e., stops) along road transects (i.e., routes) annually across North America. At each stop, observers record all birds seen or heard within a 0.4km radius during three-minute point count surveys. We summarized the data for each species at each site into $K = 5$ spatial replicates (each comprising data from 10 of the 50 stops), where each spatial replicate took value 1 if the species was detected at at least one of the 10 stops in that replicate, and value 0 if the species was not detected at any of the 5 stops. 

We focused our analysis on a community of grassland bird species following the classification of \cite{bateman2020north}. Grassland birds form one of the most threatened bird groups in North America, with many species showing steep population declines \citep{stanton2018analysis}. We restricted our analysis to include only species observed at at least 50 locations, which resulted in $N = 21$ species. As an auxiliary data source, we obtained broad-scale maps of breeding ranges for each of the 21 species from BirdLife International \citep{birdLife} and only used BBS sites that fell inside the species' range (including a 200km buffer) when estimating each individual species' effects. This resulted in substantial computational improvements, as some of the species ranges span only a small portion of the continental US (e.g., Clay-colored Sparrow (\textit{Spizella pallida})).

We obtained data on the amount of grassland habitat from the USGS EROS (Earth Resources Observation and Science) Center, which produces high-resolution (250m) annual land cover maps across the continental US that are backcasted to 1938. We calculated the average proportion of grassland within 1km of the BBS route starting location from 2010-2019. We calculated breeding season maximum temperature using data from the Parameter-elevation Regression on Independent Slopes Model (PRISM; \citealt{daly2008physiographically}) project. PRISM provides monthly, high-resolution (4km) gridded data products on minimum/maximum temperatures and precipitation across the United States. We calculated the maximum temperature between April-June at the starting location of each BBS route each year from 2010-2019, then averaged the yearly values to represent average temperature conditions during the ten-year time period. We used ten-year average covariate values rather than the values directly during the time of the point count surveys (i.e., 2019) to assess relationships to average temperature and grassland cover conditions experienced at each location (i.e., no temporal variation).

We fit a series of alternative models to assess the amount of spatial variability in the effect of maximum temperature on the 21 grassland bird species. We predicted spatial variability in the effect of maximum temperature because species-specific thermal tolerances often determine breeding range boundaries \citep{tingley2009birds}. If temperature is an important determinant of a species distribution, we would expect a greater effect of maximum temperature near the climatic extremes versus locations in the center of climate space (e.g., the greatest effect of maximum temperature would be in the coldest and hottest portions of the range; \citealt{amburgey2018range}). Further, the effect of temperature may interact with the amount of grassland area, as higher temperatures may have a stronger impact on occurrence in grass-dominated landscapes compared to landscapes with a matrix of forest and grassland \citep{jarzyna2016synergistic}. Given the rarity of some grassland species, we also sought to assess whether jointly modeling species in a multi-species framework, which allows for sharing of information across species, improved model fit and/or predictive performance. Specifically, we fit eight alternative models (Table \ref{tab:candidateModels}) to answer the following questions: (1) Is there spatial variation in the effect of maximum temperature on grassland bird occurrence?; (2) Is the spatial variation in the effect of maximum temperature adequately explained by an interaction with the amount of grassland habitat?; and (3) Do multi-species models outperform single-species models in terms of model fit and predictive performance? The candidate models all consisted of a spatially-varying intercept, but varied in whether they were single-species or multi-species, whether they included an interaction of grassland and maximum temperature, and whether they included a spatially-varying coefficient for maximum temperature (Table \ref{tab:candidateModels}). Across all models, covariates were standardized individually for each species to have mean 0 and standard deviation 1 within the given species' range. We modeled detection probability in the same manner across all eight candidate models, with a linear and quadratic effect of ordinal day of the survey and a separate intercept for each of the five spatial replicates.

We fit all models in \texttt{spOccupancy} \citep{doser2022spoccupancy}. For single-species models, we ran three chains of 200,000 iterations with a burn-in period of 100,000 and a thinning rate of 20. For multi-species models, we ran three chains of 500,000 iterations, a burn-in period of 200,000, and a thinning rate of 150. We set $q = 6$ for all multi-species models, as exploratory analysis revealed further increases in the number of factors resulted in only minimal changes in WAIC. We assessed convergence using the potential scale reduction factor (i.e., $\hat{\text{R}}$, \citep{brooks1998}), effective sample size, and visual assessment of traceplots using the \texttt{coda} package \citep{coda}. We compared the eight candidate models using WAIC as an assessment of model fit. To assess out-of-sample predictive performance, we fit the model to a random subset of 75\% of the data and subsequently predicted latent occupancy ($z_i(\bm{s}_j)$) for each species at the remaining 25\% of the locations. We compared the species-specific predicted occupancy values to a binary metric of the detection-nondetection data at each site, which took value one if the species was ever detected there across the five replicates at that location and 0 if not. We calculated the area under the receiver operating characteristic (ROC) curve (AUC; \citealt{hosmer2013applied}) using the predicted occurrence values and the hold-out data metric. AUC measures a model's discrimination power, which in this context is the ability to determine which sites are used by each species \citep{zipkin2012evaluating}. We compared predictive performance using a site-level version of the detection-nondetection data as our focus is on assessing spatial variation in occurrence of the grassland bird species. 

\subsection{Results}

\begin{figure}
    \centering
    \includegraphics[width=12cm]{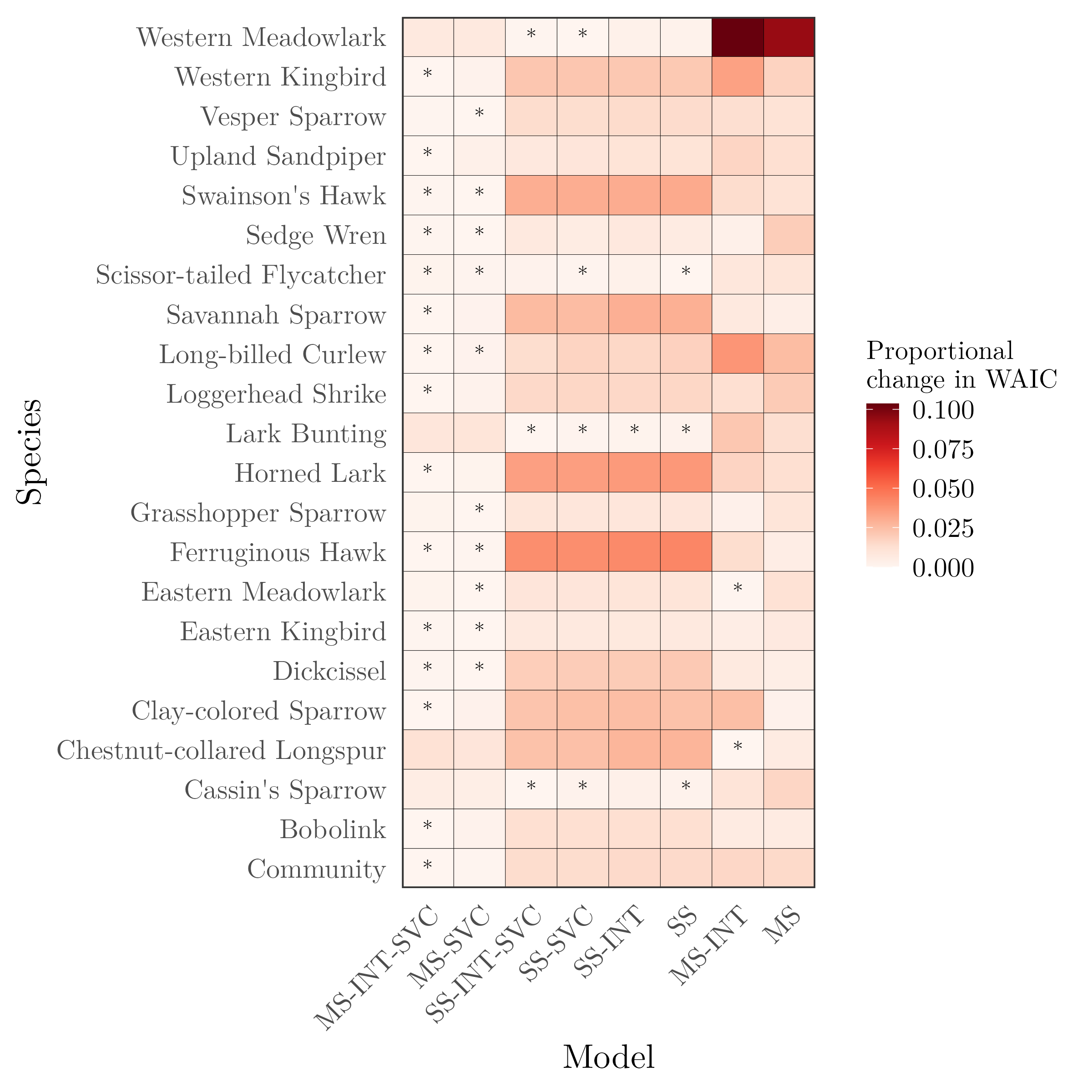}
    \caption{Model comparison results for the 21 grassland bird species across the eight candidate models using WAIC. Models differed in whether they were single-species (SS) or multi-species (MS), whether they included an interaction of grassland and maximum temperature (INT), and whether they included a spatially-varying coefficient for maximum temperature (SVC). Color corresponds to the difference in WAIC between the top model for each species and each specific model, with lighter colors indicating better model performance. Differences were divided by the minimum WAIC for each species to put all species on the same scale. The asterisk indicates the model with the lowest WAIC and those with $\Delta\text{WAIC}\geq 2$ from the top model. Note the last row corresponds to the sum across all species in the community.}
    \label{fig:waicFigure}
\end{figure}

Models with a SVC for maximum temperature substantially outperformed (i.e., $\Delta\text{WAIC} \geq 2$) models without an SVC for 16 out of 21 species (76\%) (Figure \ref{fig:waicFigure}). Models with only an interaction between maximum temperature and grassland had worse model fit compared to SVC models except for all but two species, regardless of whether SVC models also included the interaction. These results indicate that the interaction alone generally did not adequately explain spatial variation in the effect of maximum temperature. However, the model with an SVC and interaction effect had the best performance for 7 out of 21 species (33\%). Together, these results suggest the spatial variation in the effect of maximum temperature cannot be adequately explained by a simple interaction with percent grassland cover, but that this interaction may explain a portion of the spatial variation in the effect for a subset of species in the community. Modeling species jointly generally improved model fit across the community, with 17 out of 21 (81\%) species having improved fit in a multi-species model compared to a single-species model. There were less substantial differences in model out-of-sample predictive performance, as measured by AUC, for SVC models compared to non-SVC models (Supplemental Information S2). More specifically, 14 out of 21 (67\%) species had highest AUC values for SVC models, although improvements were generally minimal (Supplemental Information S2 Figure S1). Across all species, the multi-species SVC model with the interaction had the highest performance according to WAIC, and so we present the remaining results from that model.  

\begin{figure}
    \centering
    \includegraphics[width=13cm]{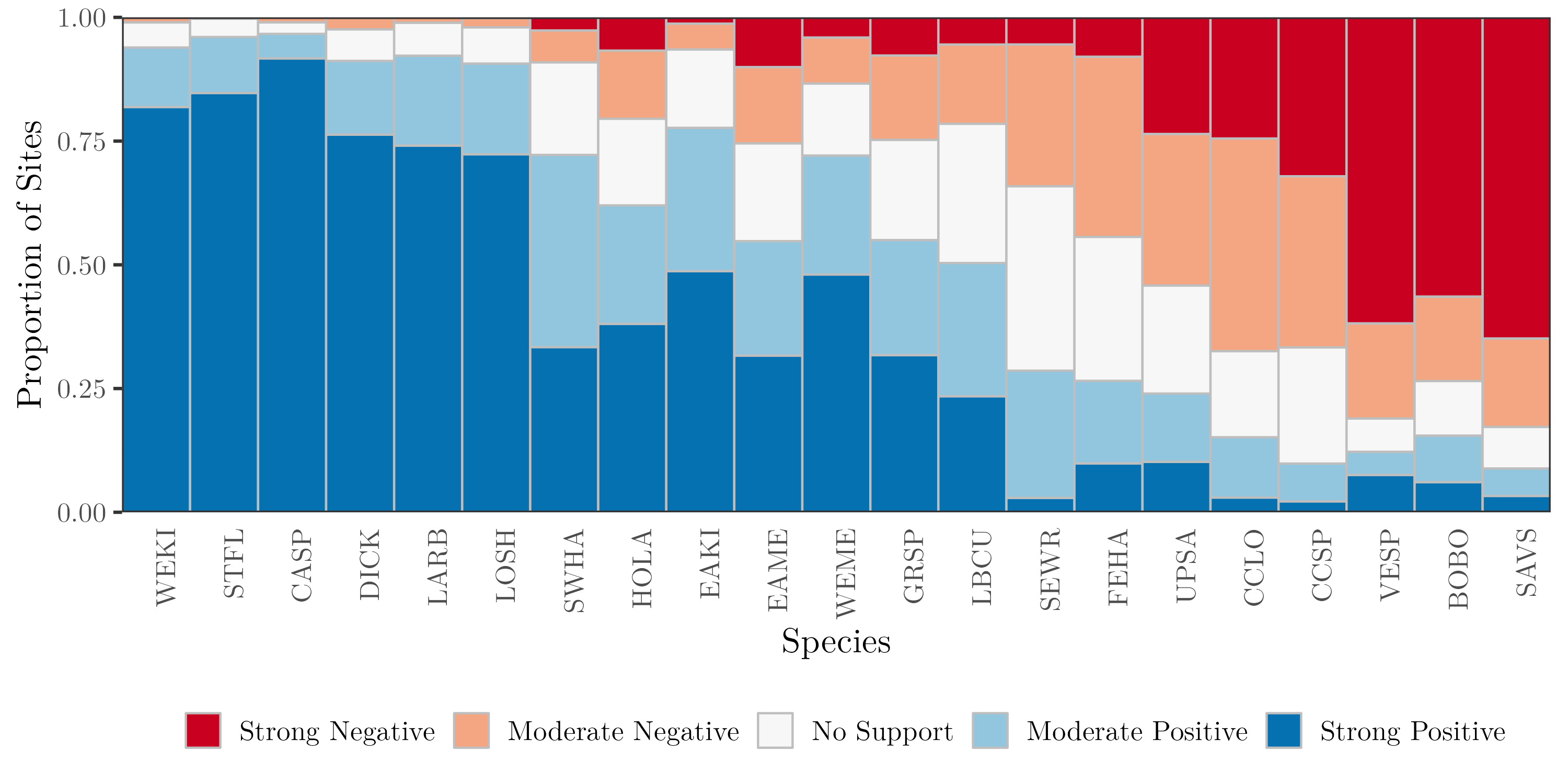}
    \caption{Overall summary of the spatially-varying effect of maximum temperature on the 21 grassland bird species. The height of each bar corresponds to the proportion of sites for the given species whose effect has the sign (i.e., positive, negative, no effect) and strength (i.e., strong, moderate) indicated by the color. Dark blue indicates strong support for positive effects and dark red indicates strong support for negative effects. More specifically: (1) Strong Positive: (P(TMAX effect $> 0$) $> 0.8$)); (2) Moderate Positive: ($0.6 < $ P(TMAX effect $> 0$) $\leq 0.8$; (3) No effect: ($0.4 < $ P(TMAX effect $> 0$) $\leq 0.6$; (4) Moderate Negative: ($0.2 < $ P(TMAX effect $> 0$) $\leq 0.4$; (5) Strong Negative: (P(TMAX effect $> 0$) $< 0.2$. See Supplemental Information S2: Table S1 for species codes.}
    \label{fig:tmax-summary}
\end{figure}

Predictions of the effect of maximum temperature across a species range revealed strong variation in the effect of maximum temperature, both within a species (i.e., across space) and among species (Figure \ref{fig:tmax-summary}). The magnitude, direction, and amount of spatial variability were highly species-specific, with some species showing large variability in the magnitude of a positive relationship with maximum temperature (i.e., WEKI (Western Kingbird)), and others showing strong support for both positive and negative relationships with maximum temperature across different portions of the species' range (i.e., FEHA (Ferruginous Hawk)). Effects of maximum temperature on ten species (SEWR, UPSA, VESP, WEME, BOBO, EAME, FEHA, GRSP, HOLA, SAVS) showed at least some degree of strong negative effects at southern portions of the range and positive effects at more northern sites (Supplemental Information S2: Figures S4-S22). This suggests that extreme breeding season temperatures and species-specific thermal tolerances are important drivers of species distributions across the community. The estimated interaction between maximum temperature and grassland cover was moderately negative on average across the community, with four species having significant negative effects (i.e., 95\% credible interval did not overlap zero; Supplemental Information S2 Figure S3). The negative interaction suggests that the effect of maximum temperature decreases as the amount of grassland increases, consistent with findings in previous work \citep{zuckerberg2018effects}. For some species (e.g., LBCU, HOLA), maximum temperature had a smaller positive effect in areas with high habitat availability (e.g., the Great Plains), while for other species (e.g., VESP), the negative interaction resulted in stronger negative effects of maximum temperature in areas with high grassland.

\begin{figure}
    \centering
    \includegraphics[width=13cm]{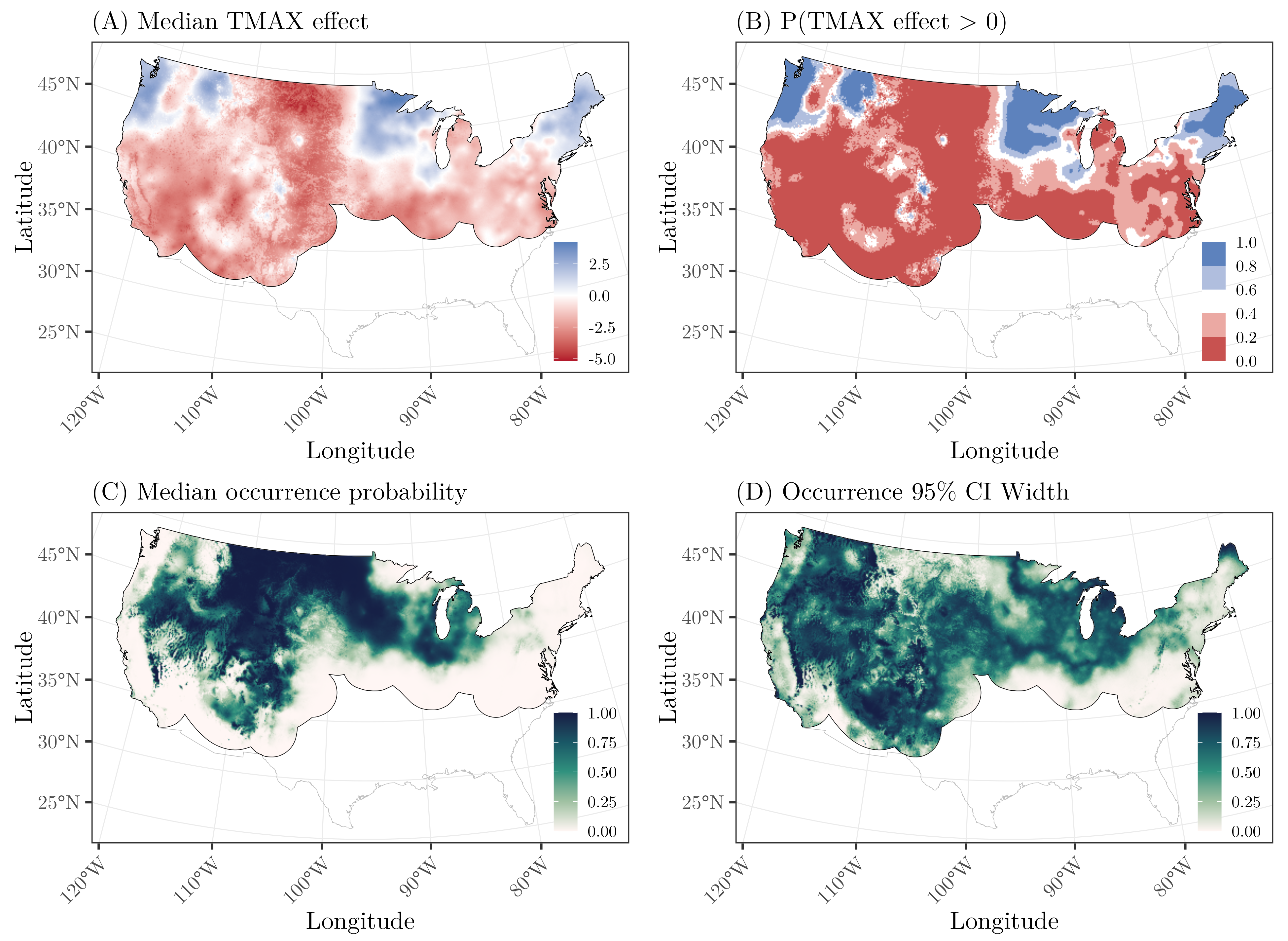}
    \caption{Posterior predictive estimates from model MS-INT-SVC of the effect of maximum temperature and occurrence probability of Vesper Sparrow (\textit{Pooecetes gramineus}). Panel (A) shows the median effect of maximum temperature, while Panel (B) shows the probability the effect is positive, where dark blue colors indicate strong support for a positive effect and dark red colors indicate strong support for a negative effect. Panel (C) shows the median occurrence probability, with associated 95\% credible interval in Panel (D).}
    \label{fig:VESP-plot}
\end{figure}

\begin{figure}
    \centering
    \includegraphics[width=13cm]{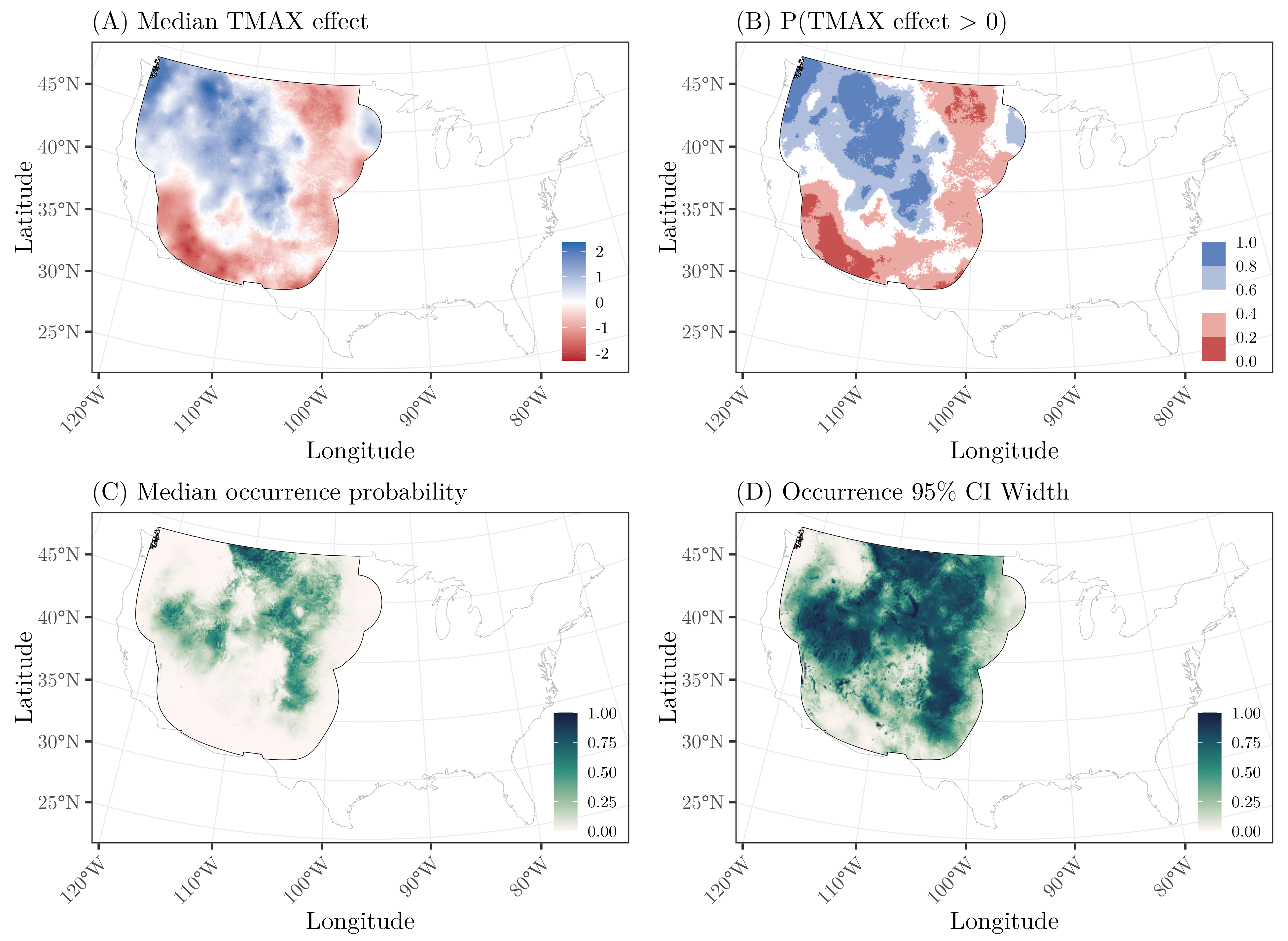}
    \caption{Posterior predictive estimates from Model MS-INT-SV of the effect of maximum temperature and occurrence probability of Ferruginous Hawk (\textit{Buteo regalis}). Panel (A) shows the median effect of maximum temperature, while Panel (B) shows the probability that effect is positive, where dark blue colors indicate strong support for a positive effect and dark red colors indicate strong support for a negative effect. Panel (C) shows the median occurrence probability, with associated 95\% credible interval in Panel (D).}
    \label{fig:FEHA-plot}
\end{figure}

Posterior predictive maps of the effect of maximum temperature can provide insight into potential distribution changes with projected increases in future maximum temperature. Figures~\ref{fig:VESP-plot} and \ref{fig:FEHA-plot} display posterior predictive maps for two example species, with the remaining 19 species shown in Supplemental Information S2. Vesper Sparrow (\textit{Poocetes gramineus}) occurrence probability was negatively related to maximum temperature throughout the southern portion of its range (Figures~\ref{fig:VESP-plot}A, B), and a significant negative interaction of maximum temperature with grassland area resulted in strongly negative effects of temperature throughout the Great Plains region, where grassland area is high (Figure \ref{fig:studyLocations}C). In much of the northern portion of the US, Vesper Sparrow occurrence probability had a positive relationship with maximum temperature. Together, these patterns suggest a potential northward shift in occurrence probability with projected increases in temperature under climate change. Ferruginous Hawk (\textit{Buteo regalis}) occurrence probability was negatively related to maximum temperature across the southern and northeastern portions of its range, while maximum temperature had a positive effect in central and western portions  of its range. The transition in the direction of the maximum temperature effect corresponds closely to the transition from the Great Plains (low elevation, warmer temperatures) to the Rocky Mountain region (high elevation, colder temperatures), suggesting Ferruginous Hawk may shift its distribution towards higher elevations.

\section{Discussion} \label{discussion}

Occupancy models are widely used to quantify species-environment relationships and predict species distributions while accounting for observational biases in data collection \citep{mackenzie2002, tyre2003improving}. Such models are increasingly applied across macroscales to predict species distributions and assess drivers of spatial and/or temporal variation in species distributions. However, existing occupancy models typically assume constant species-environment relationships; these become increasingly unrealistic as the spatial extent of analysis increases \citep{pease2022exploring}. Here, we developed computationally-efficient single- and multi-species spatially-varying coefficient (SVC) occupancy models to allow for assessment of spatially-varying species-environment relationships while explicitly accounting for imperfect detection in an occupancy modeling framework. As we demonstrated in the grassland bird case study, the use of spatially-varying coefficients in occupancy models can help elucidate the environmental factors that drive species distribution dynamics, especially across broad spatial scales. 

A key limitation for the adoption of spatially-varying coefficient models in ecology has been the substantial computational cost associated with estimating multiple Gaussian processes. Such high computational demands limit the practicality of fitting such models with commonly-used Bayesian programming languages like JAGS \citep{plummer03}, NIMBLE \citep{deValpine2017}, and Stan \citep{carpenter2017}. In the ecological literature, computationally efficient approaches for modeling SVCs include the use of stochastic partial differential equations (SPDEs; \citealt{lindgren2011explicit}) implemented within the Integrated Nested Laplace Approximation (INLA) framework \citep{rue2009approximate}, but such approaches have not been embedded in occupancy models. In the context of occupancy models, \cite{pease2022exploring} use an improper conditional autoregressive (CAR) SVC model implemented in NIMBLE to assess spatial variability in the effects of multiple covariates on four mammal species in North Carolina. Here we employed NNGPs \citep{datta2016hierarchical} within a \pg data augmentation \citep{polson2013} framework to provide an efficient implementation of Bayesian spatially-varying coefficient occupancy models that is capable of fitting models with potentially massive datasets (e.g., 1000s-100,000s of sites). NNGPs are well-defined spatial processes that provide legitimate finite-dimensional Gaussian densities with sparse precision matrices, which have been shown to perform very well compared to other Gaussian process approximations \citep{heaton2019case}. When NNGPs are embedded in an occupancy model that uses a \pg data augmentation scheme, full conditional distributions for all model parameters except the spatial decay parameters are in closed form. This reduces inefficient mixing and convergence of MCMC chains that is common when fitting Bayesian occupancy models without \pg data augmentation \citep{clark2019}, as is done by default in Bayesian programming languages like JAGS and NIMBLE. We wrote the single-species and multi-species SVC occupancy models in \texttt{C/C++} using \texttt{R}'s foreign language interface, which are available via user-friendly functions in the \texttt{spOccupancy} \texttt{R} package \citep{doser2022spoccupancy}. 

In our case study, we found substantial support for a spatially-varying effect of maximum temperature on an assemblage of 21 grassland bird species across the continental US (Figure \ref{fig:waicFigure}). The spatially-varying relationships with maximum temperature were highly variable across species (Figure~\ref{fig:tmax-summary}), which suggests individual species, despite having similar habitat requirements, may show different abilities to respond to projected increases in temperature \citep{prince2015climate}. For example, posterior predictive maps reveal Vesper Sparrow occurrence is negatively related to maximum temperature in southern portions of its range and positively related to maximum temperature in more northerly portions of the continental US (Figure~\ref{fig:VESP-plot}), suggesting a potential northward shift in the distribution of the species if temperatures continue to increase. Ferruginous Hawk occurrence is negatively related to relatively high maximum temperatures during the breeding season at lower elevations, but positively related to relatively low maximum temperature at higher elevations, indicating a potential shift towards higher elevation areas under ongoing climate change. Although speculative, these patterns demonstrate the power of SVC occupancy models to help generate hypotheses and predictions on how species may respond to future global change, which is essential information for resource managers and conservation practitioners to support management strategies. Additionally, the viability of such shifts is complicated by limited grassland and sagebrush steppe habitat in more northerly and higher elevation areas of the US, which could preclude these species from shifting their distributions to track their climatic niche. Nevertheless, the SVC predictions, and their associated uncertainties, could be used together with climate and vegetation model projections to inform identification of future climate/habitat refugia for different species assemblages, which is a critical task for ensuring species persistence under climate change \citep{saunders2023integrating}.

Our multi-species SVC occupancy model uses a spatial factor modeling approach \citep{hogan2004bayesian} for each of the spatially-varying coefficients included in the model. Such an approach is able to jointly model multiple species simultaneously, which is often shown to improve predictive performance \citep{clark2014more}. However, in our grassland bird case study, we found the multi-species SVC models performed best in terms of model fit, but did not show substantial improvements in out-of-sample predictive performance compared to single-species SVC models. These results might be related to differences in size and location of the distributions of the 21 grassland bird species, such that the spatial factor modeling approach, which assumes spatial variation in the effects of covariates can be explained by a set of common spatial factors and associated species-specific responses, may smooth over small-scale spatial variation in species-specific effects across different portions of the study region. The benefits of jointly modeling species in a multi-species SVC occupancy model compared to fitting single-species models will likely be case-specific and depend on many factors, such as the overlap of species ranges, prevalence of the species, and similarities in the different species' habitat requirements. Although not explored here, the spatial factors and species-specific factor loadings estimated in the multi-species SVC occupancy model could be used to classify species into groups based on shared spatial variation in the effects of different covariates by adapting the approaches used in \cite{doser2023joint} and \cite{shirota2019sinica} in the context of a spatially-varying intercept.  

Although \texttt{spOccupancy} harnesses the the computational efficiency provided by NNGPs and \pg data augmentation, the SVC occupancy models can still exhibit slow mixing and convergence of MCMC chains. SVC models seek to separately estimate multiple spatial processes in a single model, which can make convergence difficult to achieve, particularly for multi-species models where additional identifiability constraints need to be imposed on the matrix of factor loadings. For example, we needed to run 500,000 iterations of the multi-species SVC model in our grassland bird case study to achieve adequate convergence and mixing of the spatially-varying coefficients, which took approximately 42 hours to run using 3 cores. For multi-species models, run times can be minimized by choosing the number of factors $q$ to be as small as possible without limiting model performance. This can be determined by fitting models with varying numbers of factors, comparing models with WAIC, and choosing $q$ such that changes in WAIC become minimal as the number of factors is increased. See \cite{doser2023joint} for further discussion and recommendations for choosing the number of spatial factors. For situations where such run times are deemed untenable, similar frequentist approaches using stochastic partial differential equations may be a viable alternative, although existing software (e.g., \texttt{VAST} \citep{thorson2019guidance}) would have to be adapted to work within an occupancy modeling context. Nevertheless, while such frequentist approaches are more computationally efficient, generating uncertainty estimates for predictions of the SVCs across a region of interest is less straightforward, and direct probability statements regarding the effect of the covariate across space (e.g., Figures~\ref{fig:VESP-plot}B, \ref{fig:FEHA-plot}B) are not possible. We provide further guidance on how to improve convergence and mixing of SVC occupancy models on the package website (\url{https://www.jeffdoser.com/files/spoccupancy-web/articles/index.html}).

We envision several extensions to the SVC occupancy models presented in this paper and in the associated software implementation in \texttt{spOccupancy}. First, the models developed here estimate each SVC as arising from an independent NNGP. We could extend our approach to jointly model multiple SVCs using the multivariate NNGP SVC model described in \cite{datta2016hierarchical} to explicitly estimate correlations between the different SVCs, although this is not straightforward for the multi-species case, and it is not clear what additional gains such a model would yield. Second, we could incorporate SVCs in the detection portion of the occupancy model, which would allow effects of covariates that influence detectability (e.g., ordinal date) to vary spatially \citep{thorson2019guidance}. Lastly, we could extend SVC models to a spatio-temporal context, where we allow effects of covariates to vary spatially as well as temporally by embedding the SVCs within a dynamic linear modeling framework \citep{gelfand2005dyn}.

Spatial variability in species-environment relationships is prevalent throughout ecology \citep{rollinson2021working}. The use of spatially-varying coefficients in occupancy models can help elucidate the environmental factors that drive species distribution dynamics, especially across broad spatial scales. For accurate conclusions regarding the drivers of species distributions over space and time, we require reliable, accessible, and efficient computational methods to quantify spatially-varying relationships. We believe our proposed single-species and multi-species SVC models, and their associated user-friendly implementations in \texttt{spOccupancy}, will lead to improved inferences on the nuanced pressures facing biodiversity in a rapidly changing world. 

\section{Data Availability Statement}

All data and code are available on GitHub (\url{https://github.com/doserjef/Doser_et_al_2023_In_Review}) and will be posted on Zenodo upon acceptance. 

\section{Funding and Conflicts of Interests/Competing Interests}

We declare no conflict of interests. This work was supported by National Science Foundation (NSF) grants DBI-1954406, DMS-1916395, and DEB-2213565.

\bibliographystyle{apalike}

\bibliography{references}

\newpage

\section{Tables and Figures}

\begin{table}[ht] 
  \centering
  \caption{Model comparison results for a single-species non-spatial occupancy model (OCC), a spatially-varying intercept occupancy model (SVI), and a spatially-varying coefficients occupancy model (SVC) using simulated data with varying degrees of spatial variation in the covariate effect. Values represent the difference in four-fold cross-validation deviance and WAIC for the SVC model compared to the two candidate models. Lower values indicate better performance. Values are averaged across 50 simulated data sets.} 
  \label{tab:simStudy1}
  \begin{tabular}{|c | c | c  c  c |  c  c  c |}
    \hline
     Effective spatial & Spatial & & $\Delta$Deviance & & & $\Delta$WAIC & \\
     range (\%) & variance & OCC & SVI & SVC & OCC & SVI & SVC \\
    \hline
    10 & 0.1 & 17.30 & -0.53 & 0.00 & 25.28 & 2.54 & 0.00\\
    10 & 0.5 & 14.04 & -0.80 & 0.00 & 20.38 & 4.84 & 0.00 \\
    10 & 1.0 & 8.69 & 0.37 & 0.00 & 19.01 & 4.99 & 0.00 \\
    10 & 2.0 & 9.50 & 1.05 & 0.00 & 19.96 & 6.16 & 0.00 \\
    30 & 0.1 & 18.34 & -0.80 & 0.00 & 23.85 & 3.75 & 0.00 \\
    30 & 0.5 & 15.46 & 2.23 & 0.00 & 24.33 & 7.81 & 0.00 \\
    30 & 1.0 & 24.28 & 7.85 & 0.00 & 31.46 & 10.95 & 0.00 \\
    30 & 2.0 & 23.05 & 13.13 & 0.00 & 33.95 & 18.93 & 0.00 \\
    80 & 0.1 & 18.55 & -0.20 & 0.00 & 24.39 & 3.46 & 0.00 \\
    80 & 0.5 & 14.93 & 3.89 & 0.00 & 23.21 & 7.79 & 0.00 \\
    80 & 1.0 & 20.36 & 7.82 & 0.00 & 24.64 & 10.12 & 0.00 \\
    80 & 2.0 & 34.51 & 17.03 & 0.00 & 39.61 & 21.74 & 0.00 \\
    300 & 0.1 & 12.56 & -0.93 & 0.00 & 21.56 & 3.80 & 0.00\\
    300 & 0.5 & 14.94 & 0.28 & 0.00 & 22.28 & 4.74 & 0.00\\
    300 & 1.0 & 20.62 & 6.50 & 0.00 & 25.84 & 9.84 & 0.00\\
    300 & 2.0 & 26.26 & 11.98 & 0.00 & 31.50 & 14.77 & 0.00\\
    10,000 & 0.1 & 17.87 & -1.26 & 0.00 & 24.42 & 3.85 & 0.00\\
    10,000 & 0.5 & 14.27 & -0.40 & 0.00 & 20.46 & 3.23 & 0.00\\
    10,000 & 1.0 & 16.64 & -0.18 & 0.00 & 22.47 & 1.85 & 0.00\\
    10,000 & 2.0 & 11.51 & -0.51 & 0.00 & 17.04 & 0.88 & 0.00\\
    - & - & 13.83 & -0.59 & 0.00 & 23.44 & 3.00 & 0.00\\
    
    \hline
  \end{tabular}
\end{table}

\newpage

\begin{table}[ht] 
    \centering
    \caption{Candidate models for the grassland bird case study that represent different hypotheses regarding the effect of maximum temperature on grassland bird occurrence. Models differed in whether they were single-species (SS) or multi-species (MS), whether they included an interaction of grassland and maximum temperature (INT), and whether they included a spatially-varying coefficient for maximum temperature (SVC).}
    \label{tab:candidateModels}
    \begin{tabular}{c c c c c}
      \hline
       Model & Single-species & Multi-species & Interaction & SVC \\
      \hline
       SS & \checkmark & & & \\
       SS-INT & \checkmark & & \checkmark & \\ 
       SS-SVC & \checkmark & & & \checkmark \\
       SS-INT-SVC & \checkmark & & \checkmark & \checkmark \\
       MS & & \checkmark & & \\
       MS-INT & & \checkmark & \checkmark & \\ 
       MS-SVC & & \checkmark & & \checkmark \\
       MS-INT-SVC & & \checkmark & \checkmark & \checkmark \\
       
    \hline
    \end{tabular}
\end{table}

\newpage

\clearpage

\newpage 

\newpage

\newpage

\end{document}